# "Cultural additivity" and how the values and norms of Confucianism, Buddhism, and Taoism co-exist, interact, and influence Vietnamese society: A Bayesian analysis of long-standing folktales, using R and Stan

Quan-Hoang Vuong
Manh-Tung Ho
Viet-Phuong La
Dam Van Nhue
Bui Quang Khiem
Nghiem Phu Kien Cuong
Thu-Trang Vuong
Manh-Toan Ho
Hong-Kong T. Nguyen
Viet-Ha Nguyen
Hiep-Hung Pham
Nancy K. Napier

# "Cultural additivity" and how the values and norms of Confucianism, Buddhism, and Taoism co-exist, interact, and influence Vietnamese society: A Bayesian analysis of long-standing folktales, using R and Stan

Quan-Hoang Vuong [*,1,2] orcid, Manh-Tung Ho [1,3] orcid, Viet-Phuong La [4], Dam Van Nhue [5], Bui Quang Khiem [6], Nghiem Phu Kien Cuong [7], Thu-Trang Vuong [8] orcid, Manh-Toan Ho [1,4] orcid, Hong-Kong T. Nguyen [9] orcid, Viet-Ha Nguyen [1,9] orcid, Hiep-Hung Pham [1], Nancy K. Napier [10]

[1] Western University Hanoi, Centre for Interdisciplinary Social Research, Yen Nghia Ward, Ha Dong District, Hanoi, 100000, Vietnam

[2] Université Libre de Bruxelles, Centre Emile Bernheim, 50 av. F. D. Roosevelt, Brussels 1050, Belgium

[3] Vietnam Academy of Social Sciences, Institute of Philosophy, No.59, Lang Ha Street, Thanh Cong Ward, Ba Dinh District, Hanoi 100000, Vietnam

[4] Vuong & Associates, 3/161 Thinh Quang, Dong Da district, Hanoi 100000, Vietnam

[5] National Economics University, Giai Phong Road, Hai Ba Trung district, Hanoi 100000, Vietnam

[6] Hanoi College of Arts, 7 Hai Ba Trung street, Hoan Kiem district, Hanoi 100000, Vietnam

[7] Vietnam-Germany Hospital, 16 Phu Doan street, Hoan Kiem district, Hanoi 100000, Vietnam

[8] Sciences Po Paris, Campus de Dijon, 21000 Dijon, France

[9] Vietnam Panorama Media Monitoring, Giang Vo, Ba Dinh district, Hanoi 100000, Vietnam

[10] College of Business and Economics, Boise State University, Boise, ID 83725, USA

[*] Corresponding author; email: qvuong@ulb.ac.be

## Abstract

Every year, the Vietnamese people reportedly burned about 50,000 tons of joss papers, which took the form of not only bank notes, but iPhones, cars, clothes, even housekeepers, in hope of pleasing the dead. The practice was mistakenly attributed to traditional Buddhist teachings but originated in fact from China, which most Vietnamese were not aware of. In other aspects of life, there were many similar examples of Vietnamese so ready and comfortable with adding new norms, values, and beliefs, even contradictory ones, to their culture. This phenomenon, dubbed "cultural additivity", prompted us to study the co-existence, interaction, and influences among core values and norms of the *Three Teachings* —Confucianism, Buddhism, and Taoism—as shown through Vietnamese folktales. By applying Bayesian logistic regression, we evaluated the possibility of whether the key message of a story was dominated by a religion (dependent variables), as affected by the appearance of values and anti-values pertaining to the *Three Teachings* in the story (independent variables). Our main findings included the existence of the cultural additivity of Confucian and Taoist values. More specifically, empirical results showed that the interaction or addition of the values of Taoism and Confucianism in folktales together helped predict whether the key message of a story was about Confucianism, $\beta_{\{VT \cdot VC\}} = 0.86$. Meanwhile, there was no such statistical tendency for Buddhism. The results lead to a number of important implications. First, this showed the dominance of Confucianism because the fact that Confucian and Taoist values appeared together in a story led to the story's key message dominated by Confucianism. Thus, it presented the evidence of Confucian dominance and against liberal interpretations of the concept of the Common Roots of Three Religions (*"tam giáo đồng nguyên"*) as religious unification or unicity. Second, the concept of "cultural additivity" could help explain many interesting socio-cultural phenomena, namely the absence of religious intolerance and extremism in the Vietnamese society, outrageous cases of sophistry in education, the low productivity in creative endeavors like science and technology, the misleading branding strategy in business. We are aware that our results are only preliminary and more studies, both theoretical and empirical, must be carried out to give a full account of the explanatory reach of "cultural additivity".

**Keywords:** *Confucianism, Buddhism, Taoism, Three Religions, cultural additivity, Vietnamese culture, folktales, social norms, values, beliefs, ideals*

**JEL Classification:** A13, M14

**This manuscript version:** March 4, 2018

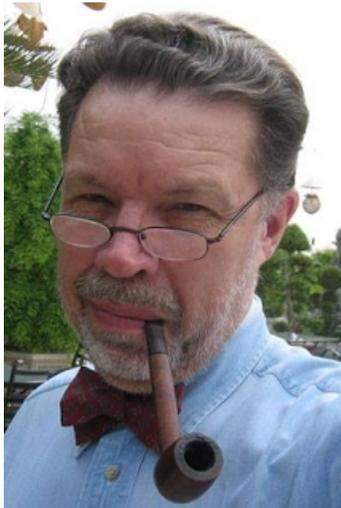

**In memory of the late professor André Farber (1943-2017)**

**"Cultural additivity" and how the values and norms of Confucianism, Buddhism, and Taoism co-exist, interact, and influence Vietnamese society: A Bayesian analysis of long-standing folktales, using R and Stan**

Quan-Hoang Vuong(*)
Manh-Tung Ho
Viet-Phuong La
Dam Van Nhue
Bui Quang Khiem
Nghiem Phu Kien Cuong
Thu-Trang Vuong
Manh-Toan Ho
Hong-Kong T. Nguyen
Viet-Ha Nguyen
Hiep-Hung Pham
Nancy K. Napier

(*) *Corresponding author*; email: qvuong@ulb.ac.be

## Introduction

*Dealing with paradox requires that one be able to hold in the mind simultaneously two diametrically opposed ideas and not go mad*. — F. Scott Fitzgerald (Aaeker & Sengupta, 2000).

The Vietnamese people reportedly burned nearly 50,000 tons of votive offerings, which include ghost money or joss papers and various paper objects, every year, according to incomplete statistics cited by its state media (Xuan, 2018). A quick calculation using the official survey of household expenditure by the Vietnamese General Statistics Office in 2016 revealed a similar trend—a household spends an average of VND 654,000/month (USD 28.8) in 2016 for ritual offerings that exclude food and fruits, up from VND 574,000/month (USD 25.2) in 2012 (GSO, 2016 Nguyen 2018).[1] By comparison, this spending was eight times as much as that for toys and children books. On the streets that sold ritual papers, one could not only find heaps of fake dollar notes but also buy paper iPhone, iPad, airplane, car, villa, to name a few paper luxury goods that are being offered to the dead. The practice has become so entrenched that every Vietnamese was familiar with it, even in favor of it. However, many either did not understand its meaning or were unaware of its Chinese origin (Lich van su, 2018). One could say that the burning of votive papers became part of Vietnamese traditions for mysterious reasons. Recently, it has become so preponderant that in February 2018, just a week after the Lunar New Year holiday (called *Tết* in Vietnamese), the Buddhist Sangha of Vietnam put forth a stunning proposal—banning the practice of burning of ritual papers at Buddhist temples altogether, due to its incompatibility with Buddhism (VNA, 2018).

---

[1] VND: Vietnamese dong, the local currency.

This kind of apparent mismatch between a traditional practice and a popular religion is in fact quite widespread in the Vietnamese culture, reflecting the local tendency to add select elements of other religions into their existing system of norms and beliefs. For instance, the façade of a house in the Old Quarters of Hanoi could comprise elements from French colonial architecture (columns and windows), Confucianism (the scroll), and Buddhism (the lotus) (as shown in figure 1). Another more revealing case is that a typical pagoda or temple here often worships both the Buddha and local deities, in addition to the ancient heroes and ancestors.

**Figure 1:** Pictures of old houses in Hanoi showing French architecture and Confucian, Taoist and Buddhist symbols were added together.

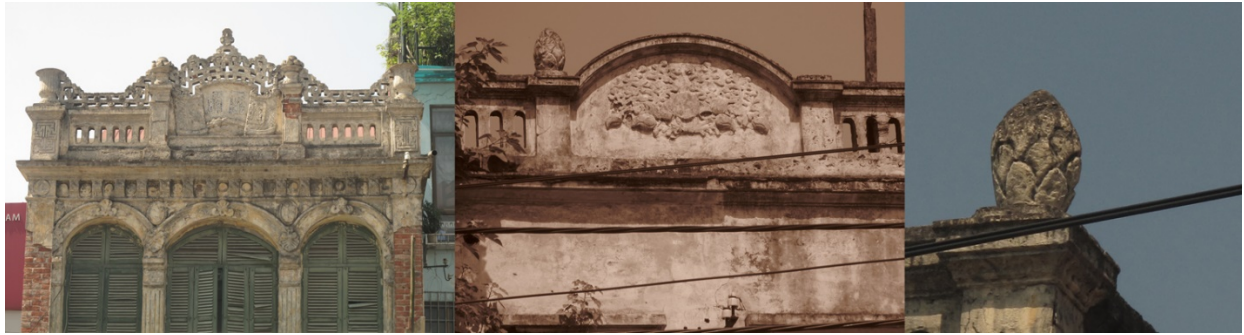

Such phenomena could be described by the concept of *cultural additivity*. The term has only been mentioned once in the academic literature, by Klug (1973). Within the scope of our analysis, the concept of cultural additivity is defined as a mechanism whereby people of a given culture are willing to incorporate into their culture the values and norms from other systems of beliefs that might or might not logically contradict with principles of their existing system of beliefs. The predominant religious philosophy and practice in Vietnam, *"tam giáo"* (literally "three teachings")—a blend of Confucianism, Taoism, and Buddhism–might be an example of this phenomenon. In other words, the background of indigenous faith was enriched over the centuries with the integration of Confucian morality, Buddhist view of the afterlife, and Taoist magical practices. Additionally, Vietnamese referred to religion as "way" (*đạo*) in common speech or "teaching" (*giáo*) in scholarly writings (Tran, 2018, p.2). There was no exact equivalent to the Latin term *religio*, which could be interpreted as either *reread* or *to go over carefully* or, more commonly, as a combination of *re-* and *ligare* (to bind) to create the meaning of *reconnection* or *relation*. The second meaning was more favored in the modern literature and also tied in with the Durkheimian conception of religion as an institution of social integration (Durkheim, 1897)—in other words, an institution that *binds* individuals together. Although religions in Vietnam did fulfill this role to an extent, the terms used to refer to religions themselves did not reflect it.

Given that the "three teachings" have been embedded quite seamlessly in the everyday life of Vietnamese people, it is no surprise that while 95% of the population (95.3 million as of July 2016) professed "religious or spiritual beliefs", only 27% consisted of religious believers, according to official data (U.S. Department of State, 2017). The report showed that more than half of the population identified as Buddhist, a fact that explained why most Vietnamese today maintain traditions originated from Buddhism-folk beliefs. At the same time, such beliefs have also given rise to superstitious rituals of unconfirmed origins, such as the burning of ghost money and goods for the dead. This phenomenon was generally regarded as a way for the living to relieve the suffering of the dead, especially those who had died in grievous circumstances (Kwon, 2007). It could also be considered as a form of communication with supernatural forces. One could observe that in the mentality of those who engaged in this type of rituals, supernatural beings were regarded as a social group that was to be negotiated with (Seth, 2013). In other words, the perpetuation of these superstitions stemmed from a confusion of society and nature.

However, this theory was insufficient in explaining the religious behaviors of the Vietnamese population, which by official account is a secular one. A complete examination must tap into the cultural and religious side, digging deeper into the roots of some beliefs, traditions and even superstitions. This study sought for the first time to bridge this gap in the literature by analyzing the association between religion and culture as shown through Vietnamese folklore.

## Literature review

### The Three Great Religions in Vietnam

#### 1. Buddhism in Vietnam

According to Nguyen (1985), Buddhism was first introduced in Vietnam as early as the second century A.D. Other scholars such as Nguyen Lang (2014) and Nguyen Tai Thu (2008, pp. 9-16), have found historical evidence that Buddhism came first-hand to Vietnam, contrary to the common belief that Vietnam acquired Buddhism second-handedly through China.

Although there were different schools of Buddhism around the world, the core beliefs and teaching of Buddhism seemed to revolve around these concepts: The Four Noble Truths, the Eight-fold Path, karma, and reincarnations. These core ideas of Buddhism have also been absorbed by the Vietnamese culture. Here we provide a brief introduction to these concepts in order to lay the context for the study.

The Buddha, when giving a diagnosis of life predicament, has come up with Four Noble Truths: i) life is *dukkha* (usually translated as "suffering" or "dissatisfaction"), ii) all sufferings are caused by *tanha* (usually translated as "desire" or "craving") iii) when one ceases to have all these desires and cravings, his/her sufferings will cease; and iv) finally, Buddhism offers the escape path out of this predicament through the Eight-fold Path: right view, right intention, right speech, right action, right livelihood, right effort, right mindfulness, and right concertation (Phillip & Smith, 2004, pp.31-49; Gethin, 1998, pp.59-79).

Furthermore, there are also many core ideas of Buddhism that have permeated the Vietnamese culture. For example, Karma (*nghiệp*) referred to the spiritual principle of cause and effect where intentions and actions of a person (cause) will affect his future outcomes (effects). Another popular concept is that of reincarnation, which is essentially determined by the karma one acquires through a lifetime. The Vietnamese also strongly believe in the idea of *"duyên"*, which is equivalent to the Sanskrit word *pratityasamutpada* and is translated into English as *"dependent origination"* or "conditionality" (Hoang, 2017).

#### 2. Confucianism in Vietnam

In the Vietnamese culture, Confucianism is more about a way of life than a religion (Nguyen, 1985). Though there might have been some Vietnamese adaptation or Vietnamization of Confucianism in Vietnam (Dong, 2015), the basic teachings on the moral codes, manners, and etiquettes for living harmoniously in a moral society remain close to the original Confucianism in the Vietnamese mind.

Vietnamese Confucianism upheld the following basic principles.

Regarding social relations or social interactions, Vietnamese people summed up the doctrines of Confucianism in *cương – thường* (纲-常). *Cương* referred to the Three Moral Bonds (*tam cương* – Sāngā; 三 纲): ruler to the ruled; parents to the children; husband to wife; in many contexts, it extended to the relationships of older siblings to younger siblings and friend to friend. *Thường* referred to the Five Cardinal Virtues (*ngũ thường – wǔ cháng* – 五常) including *Nhân* (*Rén* – 仁): benevolence or humaneness; *Nghĩa-Yì* (義): righteousness or justice; Lễ-Lǐ (禮): proper rite or propriety; Trí – Zhì (智): knowledge or wisdom; *Tín – Xìn* (信): integrity or faithfulness (Như-Quỳnh & Schafer, 1988).

Women were obliged to follow an extra set of values: Three Obediences (*tam tòng – Sāncóng –* 三从) and Four Virtues (*tứ đức – Sìdé –* 四德). The Three Obediences dictated that a woman must obey her father before marriage, obey her husband once married, and obey her son after if her husband deceased. The Four Virtues reflected not only ideal femininity but also sophisticated social qualities: moral conduct, proper speech, modest appearance, and diligent work (Taylor & Choi, 2005).

In addition, there was also another set of four virtues: *Trung – Zhong* (忠)*; Hiếu – Xiao* (孝)*; Tiết – Jie* (节)*; Nghĩa – Yi* (义)*.* The first three words prescribed the values that people of lower social rankings must uphold. *Trung* meant loyalty, as in that of a subject to his king. *Hiếu* was filial piety, obliging children to show devotion to parents. *Tiết* was usually translated to chastity or purity, which was a feminine obligation complementary to the Three Obediences; as Tran explains (1973, pp. 252 cited in Như Quỳnh & Schafer, 1988): "To venerate one's husband is *tiết*". *Nghĩa* when used together with *Tiết* in *Tiết – Nghĩa* referred to the responsibilities of the husband, in reciprocation to his wife's devotion. *Nghĩa* could also have other meanings; when appeared together with *Nhân*, *Nghĩa* could denote a principle of high moral conduct. This meaning was also well-known and could be found in common Vietnamese sayings, namely *"Respect righteousness, despise riches"* (*Trọng nghĩa khinh tài*) which encouraged people to do the right thing, not in the hope for some material reward (Như-Quỳnh & Schafer, 1988).

Confucianism was not only about social order; it also laid the foundations for institutions of governance in Vietnam, namely the mechanism of selection of imperial officials through competitive exams based on Confucian teachings. This aspect has still rung true in contemporary Vietnamese culture. For example, the educational system, business practices,  and even the law in Vietnam are often described as being heavily influenced by Confucianism (Pham, 2005; Huong & Fry, 2004; Vuong & Tran, 2009; Vuong & Napier, 2015).

## 3. Taoism in Vietnam

The earliest appearance of Taoism or Daoism in Vietnam dated back to the second century when some Taoist monks from China sought to spread their ideas to the area that is now northern Vietnam (Zhang, 2002). Zhang cited historical records as noting that Taoism in Vietnam developed the most strongly during the Chinese Tang dynasty (618-907) and later continued to exert a huge influence on the Vietnamese Lý and Trần dynasties (1010-1400).

There were two ways to interpret Taoism: the philosophical Taoism *(Đạo gia)* and the religious Taoism *(Đạo giáo)*. Based on the classics *Daodejing* of Zhuangzi and Huainaizi, the philosophical Taoism offered a worldview based on the natural approach to life. Vietnamese people were introduced to this philosophy through concepts such as dao, the yin and yang, the Five elements, the ethics of "noncontrivance" or "effortless action" (*vô vi*) (Slingerland, 2007) and "spontaneity" (*tự nhiên*). The way of life in philosophical Taoism was largely aimed towards reaching the *wuwei* (无为) realm. The word *wuwei* meant doing nothing in the sense of letting life flow naturally. Nature was the leitmotif of the Taoist philosophy and truly set Taoism apart from Confucianism (Novak, 1987). Consequently, the image ideal of a person who retreats to the nature and free from the constricted life of Confucian rules is usually associated with Taoism.

Religious Taoism, like Buddhism, did not have a core system of specific teachings. Broadly speaking, its practice focused on the search for longevity and immortality, spiritual healing practices, magics, and divinations, which blended in with Vietnamese popular religious beliefs (Tran, 2017, pp. 13). Unlike in China, Taoism in Vietnam took no institutional form, in the sense that there were no Taoist schools. Practitioners of Taoism, called "*masters*" (*thầy*) were often shaman-like specialists in a variety of domains such as healing, ritual sacrifice (at funerals, for example), soothsaying, sorcery, geomancy, etc., and were often not attached to any temple. In fact, Taoist temples did not serve the role of training monks and priest; rather, they were places of worship for immortals (historical figures who had been 'canonized' in Vietnamese culture and folk beliefs) and Taoist deities such as the Jade Emperor (*Ngọc Hoàng*). The Vietnamese Taoist pantheon was widely accepted by the population, to the point that they weren't recognized as Taoist deities anymore, rather simply considered as traditionally worshipped gods.

### *"Tam giáo đồng nguyên"*– Three Religions of Common Roots in Vietnam

The three teachings (*Tam giáo*) of Confucianism, Buddhism, and Taoism did not gain its foothold among Vietnamese population in a vacuum. Like all other countries in the world, Vietnam had their own popular religion, beliefs that are held by educated classes as well as unschooled commoners alike (Cleary, 1991). Similar to popular religion in China, the Vietnamese popular religion was characterized by the worshipping of ancestors, local deities and goddess, local festivals honoring local gods (especially village gods), various forms of exorcism of harmful forces, spirit-possession, the practices of divinations; the offering to deities, goddess, even the Buddha for luck (in the forms of good weather, good harvest, children, health, etc.) (Lizhu, 2003; Toan Anh, 2005, pp. 13). It was also common for people to worship historical heroes, sometimes even people who have accidental deaths (Tran Q. Anh, 2018, pp. 3).

Against this background, the first official examination based on the Three Religions was given during the prime of Lý Dynasty, in 1195, 1227 and 1247 (Phan Huy Chu, 2014). The Lý-Trần dynasties had shaped a principle of relative religious pluralism, originating from China called *"Tam giáo đồng nguyên"* (三教同源). There have been so many ways in which scholars in the field of religion and culture seek to define the meaning of this concept.

Le (2016) rendered the concept as *Common Root of Three Religions*, which could refer to not only common origins but also the co-existence, the convergence, and even the unification of the three religions Buddhism, Confucianism, and Taoism in Vietnam. While this may suggest an equal footing between these three major religions of Vietnam, other scholars have argued Taoism is, in fact, played a lesser role in terms of influence on society and politics (Xu, 2002; Nguyen, 2015). Nguyen (2015) also pointed out that Buddhism catered to the spiritual aspect of the population while Confucianism served as a base for a political and institutional organization. With these two main aspects of society taken care of, Taoism naturally faded more to the background on the religious scene, especially considering how the philosophy of Taoism aimed to avoid conflicts and the ideal Taoist would retreat to isolation in nature rather race for influence.

How do the values and norms of Confucianism, Buddhism, and Taoism co-exist, interact and influence each other in Vietnamese culture? We set out to answer this question by looking at the values and norms of these faiths as expressed in Vietnamese popular folktales.

## Materials and Data Construction

### Materials

As people tell each other stories in order to reinforce social and cultural norms and values, the oral nature of the myths and tales was an important criterion for the purpose of this study, which was to reflect on the cultural values in popular Vietnamese folktales throughout history. Thus, we selected only old tales and stories that survive in the contemporary narrative after centuries of oral transmission.

Storytelling, which dated back to ancient times and is defined as the oral interpretation of stories with its own flavor and tradition, was well-known for its influence as a medium for the transmission of ideas and culture (Baker & Green 1987). Schiro pointed out that prior to the invention of writing, early storytellers, who were historians, entertainers, news carriers, priests, artisans, and teachers, were the medium through which a society passed on its culture. Writing, especially once the cost of printing went down in the fifteenth century, gave the educated a new medium to pass on their culture (Schiro 2004). Therefore, to understand the additivity of various elements within the Vietnamese culture, it was necessary to examine the texts that have thrived in the verbal and written forms in the Vietnamese folklore of the past and today.

The following sources had been used for data gathering:

- 88% of the total materials are the collection of Vietnamese folktales: (i) either collected and rewritten by Nguyễn Đổng Chi (2014a, 2014b) in *Kho tàng truyện cổ tích Việt Nam* (*Collection of Vietnamese folktales*, originally published in 1957) (ii) or ancient folktales posted online. The work of Nguyễn Đổng Chi was believed to be one of the most systematic folktale collections as it touched on the characteristics, origins, and historical development of Vietnamese folklore as well as provided variants of folkloric motifs in the folklore of other countries and ethnicities (On, 2016). Permeating through these stories, which were all in prose narrative with a sprinkle of poetry, were the themes of folk history and mythology, manifested in the desires of the people to have enough to eat, to sleep sound, to have a benevolent ruler, and to be blessed by the deities and supernatural beings.

There were also four major collections of stories known in the Vietnamese language as *truyền kỳ*, a term that originated from the Chinese word 传奇 (narration of strange things):

- The fourteenth-century collection of tales written in Chinese and compiled by Lý Tế Xuyên called *"Việt Điện U Linh"* (Spiritual powers in the Viet realm) (Xuyen, 2013).

- The fifteenth-century compilation of stories "*Lĩnh Nam Chích Quái*" (Selected tales of extraordinary beings in Linh Nam) (Phap, 2017).
- The sixteenth-century collection of strange events "*Truyền Kỳ Mạn Lục*" (Giant anthology of strange tales) by Nguyễn Dữ (Du, 2016). This book consisted of 20 stories, all written in Chinese script and later translated into the ancient Vietnamese script (*Nôm*), and then today's Vietnamese national language today (*Quốc ngữ*). This study used the translation by Trúc Khê in 1943 as it seemed to be the most popular version to date. Stories in this collection took place in the northern region of Vietnam and spanned a wide timeline, either during the Lý (1010-1225), the Trần (1225-1400), the Hồ (1400-1470) or the first half of the Later Lê dynasties (1428-1527). This was because the author lived in the sixteenth century under the Lê period—when the influence of contemporary Chinese Neo-Confucianism in Vietnam was at its height (Whitmore, 1984).
- "*Thánh Tông di thảo*" (*The remaining drafts of King Thánh Tông*) was commonly believed to be the work of Le Thanh Tong (1460 – 1497) – a famous king in the Later Lê dynasty of Vietnam. The book contains 19 anthological, fabled and memoir stories (Tong, 2017).

Within the first months of 2018, we have collected 345 lines of data as the results of reading 307 different stories from the books mentioned above. All the data are presented openly at *Open Science Framework* [DOI: http://doi.org/10.17605/OSF.IO/8F7RM; ARK: c7605/osf.io/8f7rm, folder "Estimations"] according to the principle of open data (Vuong, 2017).

**Construction of Variables**

Data team members carefully read and critically assessed individual story, before putting a value of 1 or 0 for each variable according to the rules described in what follows. As a general rule, each story corresponded to one data line, when the ending of the story was predominantly about one character or a group of similar characters. However, a story may have a protagonist and an antagonist with very different endings; in this case, it could result in two lines of data input.

***Dependent variables – the categorization of the stories according to the Three Great Religions:***

- **C – short for Confucianism:** value **1**: the story sent a message about Confucianism and/or Confucian values; value **0**: the aforementioned values were absent. Here, Confucian values are understood through the three types of relationship: Ruler-Ruled, Father-Son, and Husband-Wife, in addition to the five virtues: Nhân – Rén (仁): benevolence or humaneness; Nghĩa-Yì (義): righteousness or justice; Lễ – Lǐ (禮): proper rite or propriety; Trí – Zhì (智): knowledge or wisdom; Tín – Xìn (信): integrity or faithfulness. Refer to section *Confucianism in Vietnam* in *The Three Great Religions in Vietnam* in *Literature Review* above for more details.
- **T – short for Taoism:** value **1**: the story was centered around magical cultivation and practices, ways to attain the immortal life as well as reach the *wuwei* realms; value **0**: no such message was delivered. This category concerned Taoism and Taoist practices; thus, it was reserved for any stories that reflected the commoners' wish to live a life with magical power and free from fears and worries. Refer to section *Taoism in Vietnam* in *The Three Great Religions in Vietnam* in *Literature Review* above for more details.
- **B – short for Buddhism:** value **1**: the story concerned self-cultivation according to Buddhist principles; featured characters from Buddhist teachings (including Buddha who often made an appearance in Vietnamese folktales in the same manner as fairy godmothers did in European fairy tales); sent a moral message about escaping the cycle of suffering, reincarnation, and karma; value **0**: none of the above was mentioned. Refer to section *Buddhism in Vietnam* in *The Three Great Religions in Vietnam* in *Literature Review* above for more details.

***Independent variables – Attitudes and behaviors of the main characters within a story***

The independent variables were established as follows:

- **VB – or values of Buddhism**: value **1**: the character behaved in accordance with the core values of Buddhism (e.g. praying to the Buddha or obeying the instruction of a bodhisattva); value **0**: the character did not behave as such.
- **VT – or values of Taoism**: value **1**: the character behaved in accordance with the core values of Taoism (e.g. practicing or believing in Taoist magic); value **0**: the character did not behave as such.
- **VC – or values of Confucianism**: value **1**: the character behaved in accordance with the core values of Confucianism (e.g. filial piety, fidelity, and other Confucian values as listed above); value **0**: the character did not behave as such.
- **AVB – anti-Buddhist-values**: value **1**: the main character behaved against the core values of Buddhism (e.g. murder, religious blasphemy, overly indulged with sex); value **0:** the main character did not behave as such.
- **AVT – anti-Taoist-values:** value **1**: the main character behaved against the core values of Taoism (e.g. abandoning the wuwei path); value **0**: the main character did not behave as such.
- **AVC- anti-Confucian-values**: value **1**: the main character behaved against the core values of Confucianism (e.g. disobeying the order of the rulers, lacking filial piety toward parents, living in discord between husband and wife); value **0:** the main character did not behave as such.

## Methods of analysis

Inspired by the principle of additivity in probability as well as the observed phenomenon of Vietnamese people arbitrarily selecting and adding beliefs – that might sometimes appear contradictory to principles of their existing beliefs – to their culture, we have come up with the following categorical logit model.

### Conceptualization of the model

The diagram in Figure 2 visually presented the logic of the model. When approaching a Vietnamese folktale, one should take into account the co-existence of Confucianism, Taoism and Buddhism, and the possible interaction in influencing folk mentality.

**Figure 2: Three models of cultural additivity of religious faiths (a, b,c)**

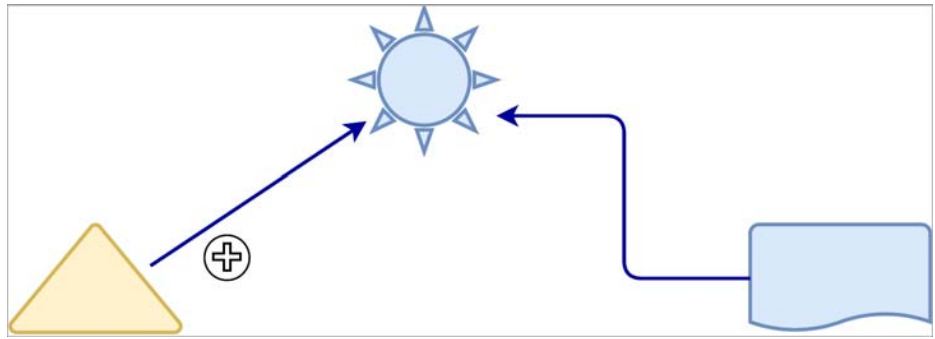

a) Simple reflection with no cultural additivity

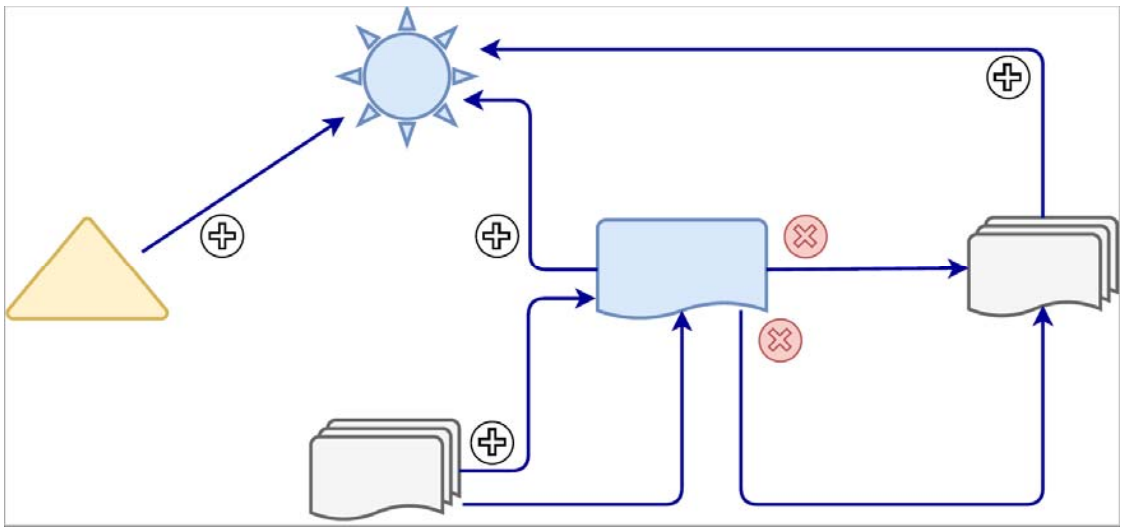

b) Complex cultural additivity

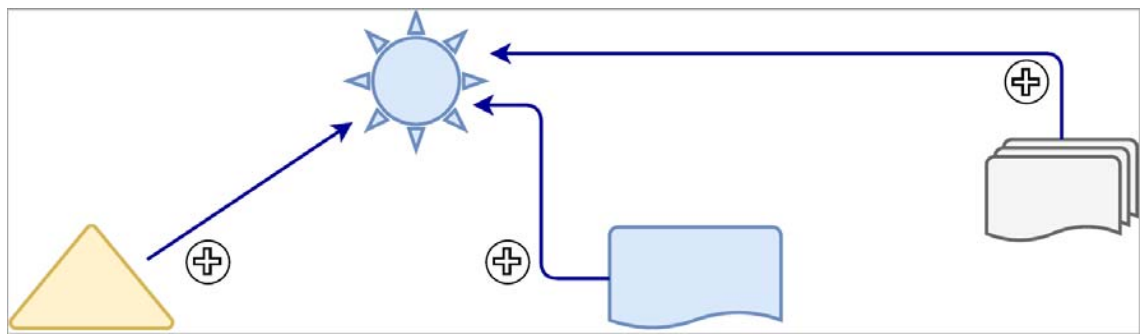

c) Simple cultural additivity

(*) Notes for Figure 2 (Three models of cultural additivity of religious faiths): The "sun" represents the nature of the key message of the stories: whether it is about Confucianism, Buddhism or Taoism (the dependent variable). The triangle represented the appearance of anti-values of whichever religion the sun represents. The light blue folder represented the appearance of the values of the concerned religion. The white folder represented the appearance of values of other religions.

Figure 2a is a simple model with no cultural additivity: the story conveyed a key message dominated by one religion. This fact was predicted through the appearance of the values and anti-values of said religion. Figure 2b was a complex model of cultural additivity. In this model, whether the key message a story conveys was dominated by a religion was predicted through the appearance of its values multiplied by the probability of appearance of the values and anti-values of other religions. Figure 2c was a simple model of cultural additivity: the probability that the key message conveyed by a story was dominated by a religion was determined by the appearance of its values and anti-values, as well as the appearance of the values of other religions.

According to this conceptualization, we could construct the categorical logistic equation to predict whether a story conveys a key message about a religion through these models of cultural additivity.

**Methods: Estimation models**

As all variables are dichotomous and the data for response and predictor variables are discrete, we employed logistic models. Logistic models are used to predict the probability of each value of the dependent variable given specific values of the independent variables (see McElreath (2016), Chapter 10, for detailed technical explanations). They are presented in the series below.

Models series for Buddhism or B contains equations (mB1-3):

$$
\begin{aligned}
&B \sim \text{Binomial}(1, p) \quad &\text{(mB1)}\\
&\text{logit}(p) = \alpha_B + \beta_{\{VB\}} \cdot VB + \beta_{\{AVB\}} \cdot AVB\\
&\alpha_B \sim \text{Normal}(\mu, \sigma)\\
&\beta_{\{VB\}} \sim \text{Normal}(\mu, \sigma)\\
&\beta_{\{VAB\}} \sim \text{Normal}(\mu, \sigma)
\end{aligned}
$$

$$
\begin{aligned}
&B \sim \text{Binomial}(1, p) \quad &\text{(mB2)}\\
&\text{logit}(p) = \alpha_B + \left(\beta_{\{VB\}} + \beta_{\{VC\}} \cdot VC + \beta_{\{VT\}} \cdot VT\right) \cdot VB + \beta_{\{AVB\}} \cdot AVB\\
&\alpha_B \sim \text{Normal}(\mu, \sigma)\\
&\beta_{\{VB\}} \sim \text{Normal}(\mu, \sigma)\\
&\beta_{\{VC\}} \sim \text{Normal}(\mu, \sigma)\\
&\beta_{\{VT\}} \sim \text{Normal}(\mu, \sigma)\\
&\beta_{\{VAB\}} \sim \text{Normal}(\mu, \sigma)
\end{aligned}
$$

$$
\begin{aligned}
&B \sim \text{Binomial}(1, p) \quad &\text{(mB3)}\\
&\text{logit}(p) = \alpha_B + \left(\beta_{\{VB\}} \cdot VB + \beta_{\{VC\}} \cdot VC + \beta_{\{VT\}} \cdot VT\right) + \beta_{\{AVB\}} \cdot AVB\\
&\alpha_B \sim \text{Normal}(\mu, \sigma)\\
&\beta_{\{VB\}} \sim \text{Normal}(\mu, \sigma)\\
&\beta_{\{VC\}} \sim \text{Normal}(\mu, \sigma)\\
&\beta_{\{VT\}} \sim \text{Normal}(\mu, \sigma)\\
&\beta_{\{VAB\}} \sim \text{Normal}(\mu, \sigma)
\end{aligned}
$$

Models series for Confucianism or C contains equations (mC1-3):

$$
\begin{aligned}
&C \sim \text{Binomial}(1, p) \quad &\text{(mC1)}\\
&\text{logit}(p) = \alpha_C + \beta_{\{VC\}} \cdot VC + \beta_{\{AVC\}} \cdot AVC\\
&\alpha_C \sim \text{Normal}(\mu, \sigma)\\
&\beta_{\{VC\}} \sim \text{Normal}(\mu, \sigma)\\
&\beta_{\{VAC\}} \sim \text{Normal}(\mu, \sigma)
\end{aligned}
$$

$$C \sim \text{Binomial}(1, p) \hspace{3cm} \text{(mC2)}$$
$$\text{logit}(p) = \alpha_C + (\beta_{\{VC\}} + \beta_{\{VB\}} \cdot VB + \beta_{\{VT\}} \cdot VT) \cdot VC + \beta_{\{AVB\}} \cdot AVC$$
$$\alpha_C \sim \text{Normal}(\mu, \sigma)$$
$$\beta_{\{VB\}} \sim \text{Normal}(\mu, \sigma)$$
$$\beta_{\{VC\}} \sim \text{Normal}(\mu, \sigma)$$
$$\beta_{\{VT\}} \sim \text{Normal}(\mu, \sigma)$$
$$\beta_{\{VAC\}} \sim \text{Normal}(\mu, \sigma)$$

$$C \sim \text{Binomial}(1, p) \hspace{3cm} \text{(mC3)}$$
$$\text{logit}(p) = \alpha_C + (\beta_{\{VB\}} \cdot VB + \beta_{\{VC\}} \cdot VC + \beta_{\{VT\}} \cdot VT) + \beta_{\{AVC\}} \cdot AVC$$
$$\alpha_C \sim \text{Normal}(\mu, \sigma)$$
$$\beta_{\{VB\}} \sim \text{Normal}(\mu, \sigma)$$
$$\beta_{\{VC\}} \sim \text{Normal}(\mu, \sigma)$$
$$\beta_{\{VT\}} \sim \text{Normal}(\mu, \sigma)$$
$$\beta_{\{VAC\}} \sim \text{Normal}(\mu, \sigma)$$

Models series for Taoism or T contains equations (mT1-3):

$$T \sim \text{Binomial}(1, p) \hspace{3cm} \text{(mT1)}$$
$$\text{logit}(p) = \alpha_T + \beta_{\{VT\}} \cdot VT + \beta_{\{AVT\}} \cdot AVT$$
$$\alpha_T \sim \text{Normal}(\mu, \sigma)$$
$$\beta_{\{VC\}} \sim \text{Normal}(\mu, \sigma)$$
$$\beta_{\{AVT\}} \sim \text{Normal}(\mu, \sigma)$$

$$T \sim \text{Binomial}(1, p) \hspace{3cm} \text{(mT2)}$$
$$\text{logit}(p) = \alpha_T + (\beta_{\{VT\}} + \beta_{\{VB\}} \cdot VB + \beta_{\{VC\}} \cdot VC) \cdot VT + \beta_{\{AVB\}} \cdot AVT$$
$$\alpha_T \sim \text{Normal}(\mu, \sigma)$$
$$\beta_{\{VB\}} \sim \text{Normal}(\mu, \sigma)$$
$$\beta_{\{VC\}} \sim \text{Normal}(\mu, \sigma)$$
$$\beta_{\{VT\}} \sim \text{Normal}(\mu, \sigma)$$
$$\beta_{\{AVT\}} \sim \text{Normal}(\mu, \sigma)$$

$$T \sim \text{Binomial}(1, p) \hspace{3cm} \text{(mT3)}$$
$$\text{logit}(p) = \alpha_T + (\beta_{\{VT\}} \cdot VT + \beta_{\{VB\}} \cdot VB + \beta_{\{VC\}} \cdot VC) + \beta_{\{AVB\}} \cdot AVT$$
$$\alpha_T \sim \text{Normal}(\mu, \sigma)$$
$$\beta_{\{VB\}} \sim \text{Normal}(\mu, \sigma)$$
$$\beta_{\{VC\}} \sim \text{Normal}(\mu, \sigma)$$
$$\beta_{\{VT\}} \sim \text{Normal}(\mu, \sigma)$$
$$\beta_{\{AVT\}} \sim \text{Normal}(\mu, \sigma)$$

The categorical regression for the dichotomous predicted and predictor variables was carried out thanks to the computing power of Stan's Hamiltonian MCMC in the statistical software R ( release 3.3.3). We performed this regression both directly using Stan codes for files named mB1.stan, mB2.stan, mB3.stan, mC1.stan, mC1.stan, mC2.stan, mC3.stan, mT1.stan, mT2.stan and mT3.stan.

The files were deposited with the open code principles at *Open Science Framework* [DOI: http://doi.org/10.17605/OSF.IO/8F7RM; ARK: c7605/osf.io/8f7rm, folder "Estimations"]. The box below presented an example of Stan codes for estimation of mC2.stan:

```
data{
    int<lower=1> N;
    int C[N];
    int VB[N];
    int VT[N];
    int VC[N];
    int AVC[N];
}
parameters{
    real a;
    real bVC;
    real bVB;
    real bVT;
    real bAVC;
}
model {
    vector[N] lp;
    bAVC ~ normal ( 0 , 10 );
    bVT ~ normal ( 0 , 10 );
    bVB ~ normal ( 0 , 10 );
    bVC ~ normal ( 0 , 10 );
    a ~ normal ( 0 , 10 );
    for ( i in 1:N ) {
        lp[i] = a + (bVC + bVB * VB[i] + bVT * VT[i]) * VC[i] + bAVC * AVC[i];
    }
    C ~ binomial_logit( 1 , lp );
}
generated quantities{
    vector[N] lp;
    real dev;
    dev = 0;
    for ( i in 1:N ) {
        lp[i] = a + (bVC + bVB * VB[i] + bVT * VT[i]) * VC[i] + bAVC * AVC[i];
    }
    dev = dev + (-2)*binomial_logit_lpmf( C | 1 , lp );
}
```

Another example of R commands to run Stan codes was deposited in the R file "est_stan_180226_v5.R". The computer codes in this file employed a combination of utility programs called from the R file: "DBDA2E-utilities.R", programs use with the book *Doing Bayesian Data Analysis, Second Edition* (Krusche, 2015).

The results of estimations using the above codes were also tested with the "Rethinking Package" v1.59 of Richard McElreath (URL: https://github.com/rmcelreath/rethinking). Other programs to run estimations for each model in the afore-mentioned series were packaged in the R file: "20180226b_rethinking_map.R". The estimation had in total 8000 iterations, divided into 4 Markov chains, each containing 1000 warmup iterations.

## Estimations results

### Technical validation

The results of initial estimations for the model (mC2) are presented in figure 3 as follows. In general, the results satisfied the conditions of standard distribution of the model's posteriors.

**Figure 3:** Model (mC2)'s posterior distribution satisfying standard distributions.

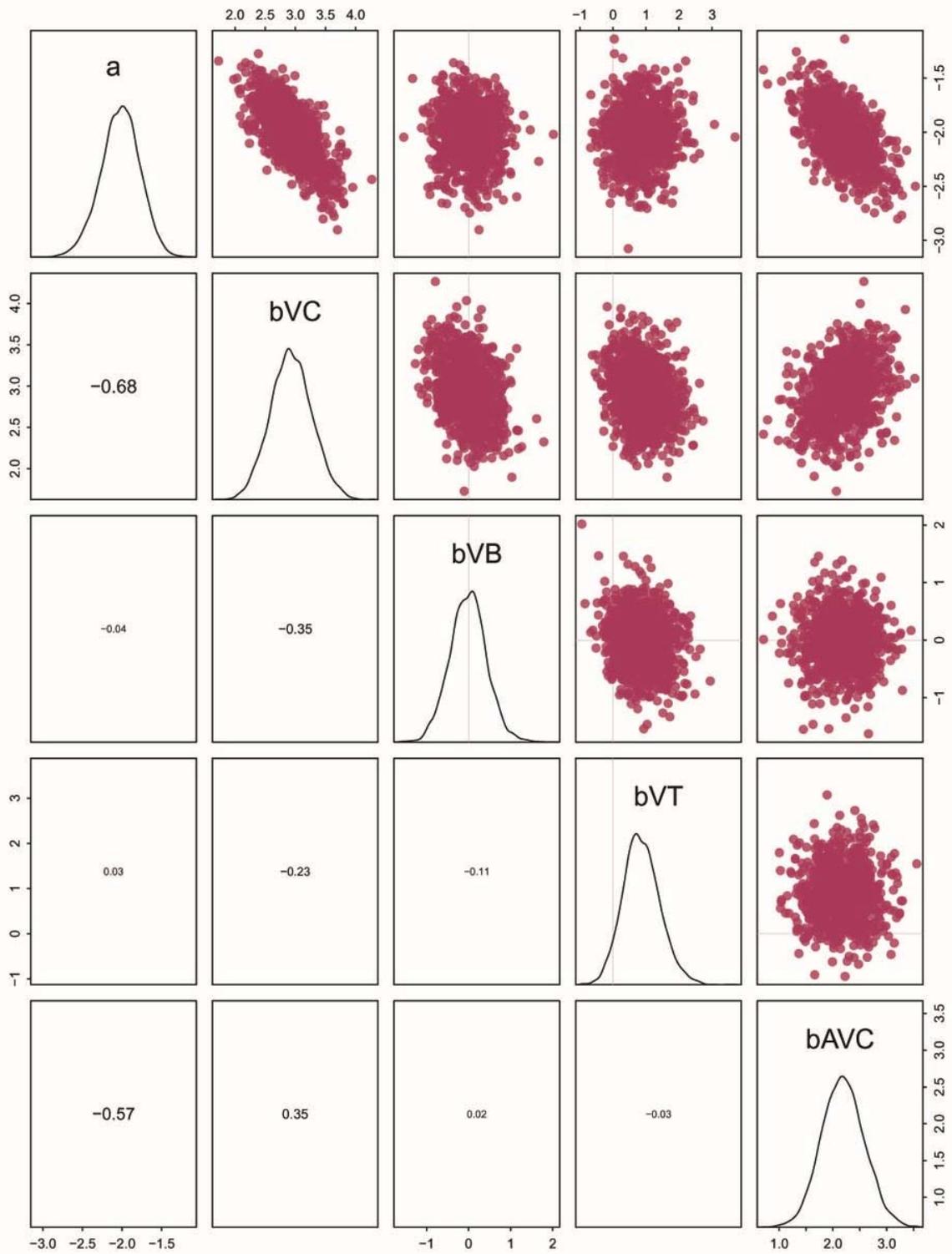

(*)Note for Figure 3: Model (mC2) represents the complex additivity model of the appearance of Buddhist and Taoist values and the predictor for Confucianist key message of a folktale.

The same observation could be made from running estimations for other models of Confucianism (mC1, mC3) and the models of Buddhism (mB1, mB2, mB3) and Taoism (mT1, mT2, mT3). All of the charts could be viewed and downloaded from *Open Science Framework*, DOI: 10.17605/OSF.IO/8F7RM; ARK: c7605/osf.io/8f7rm, folder "Suppl", which contains supplementary files.

From the estimations, one could report the basic statistics of posteriors. Figure 4 presented the posterior distributions of the coefficients, together with the mean, mode and median values from the model (mB3), i.e. the model of simple additivity for Buddhist values.

The distributions given in Figure 4 also show the high probability density regions for the evaluated coefficients, with the default probability of 95%. This probability has nothing to do with the frequentist 5% "conventional level of significance"; it is just the habit of looking into normal distributions. (See Kruschke (2015) and McElreath (2016) for a full account of technical details regarding Bayesian statistics versus frequentist practices.)

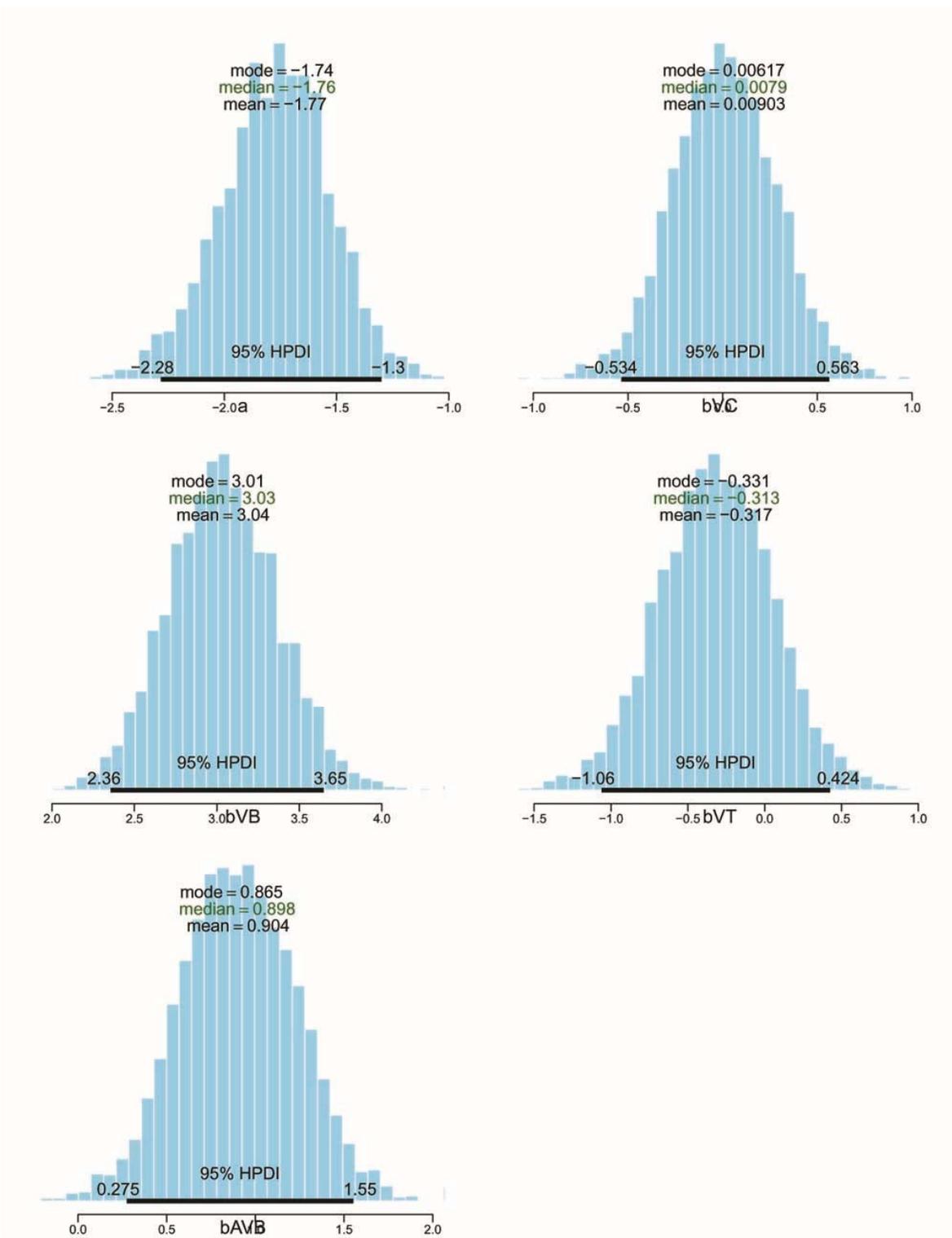

**Figure 4:** Posterior distributions of all coefficients from the model (mB3), the model of simple additivity for the appearance of Buddhist and Taoist values as the predictors for the Buddhist key massages in a folktale.

In figure 5 below, we presented an example of testing the validity of the coefficients $\beta_{\{VC\}}$ in the model (mC3) for simple additivity of Confucian values. Here, one can see the effective sample size of the chains was reasonably large, ESS > 2300. That means there were nearly 60% of total posterior calculation points after warmup, showing satisfactory computational efficiency.

In addition, the chains fluctuated around $\beta_{\{VC\}} = 3.11$, and were well-mixed (Figure 5, top left). The results of running the autocorrelation function also show technical validity, it fluctuated around $\text{coeff}_{\{AC\}} = 0$ after the lag of 3, for all four chains. Shrinking factor of computed mean values converged to 1.0 quite fast.

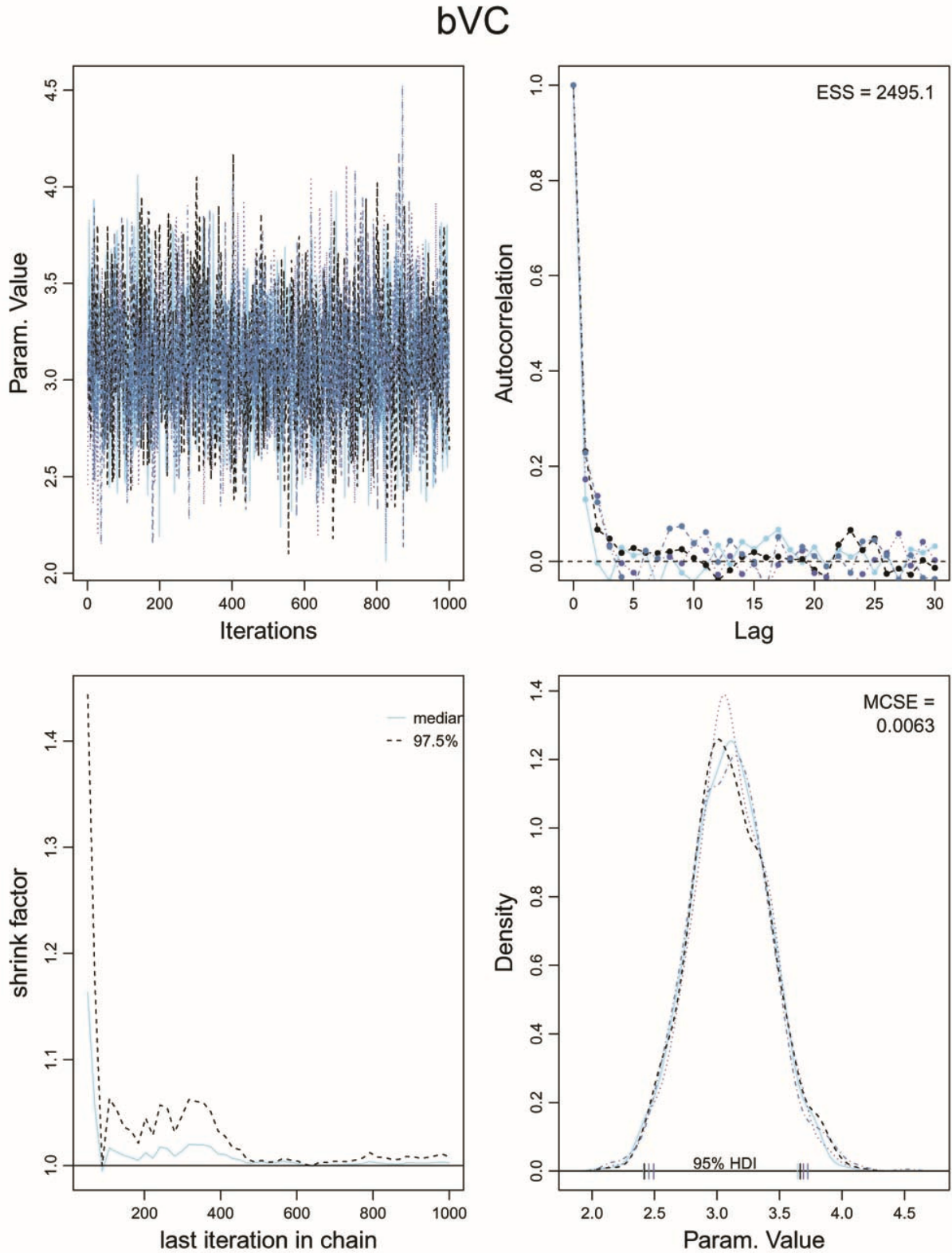

**Figure 5:** Technical validation of the coefficients $\beta_{\{VC\}}$ in the model (mC3), the model for simple additivity of Confucianist and Taoist values as the predictor for the Confucian key message in a folktale.

**Checking the robustness of the evaluations**

In order to check the robustness of the models through prior and to create a visual presentation of the test, we have used Rethinking Package in R. The following box presented the codes.

```
mC.3 <- map2stan(
alist(
C ~ dbinom( 1 , Ip ) ,
logit(Ip) <- a + (bVC*VC + bVB*VB +bVT*VT) + bAVC*AVC,
a ~ dnorm(0, 10),
bVC ~ dnorm(0, 10),
bVB ~ dnorm(0, 10),
bVT ~ dnorm(0, 10),
bAVC~ dnorm(0, 10)
) ,
chains = 4,
data=dat )

mC.3a <- map2stan(
alist(
C ~ dbinom( 1 , Ip ) ,
logit(Ip) <- a + (bVC*VC + bVB*VB +bVT*VT) + bAVC*AVC,
a ~ dnorm(0, 10),
bVC ~ dnorm(-3, 1),
bVB ~ dnorm(0, 10),
bVT ~ dnorm(0, 10),
bAVC~ dnorm(0, 10)
) ,
chains = 4,
data=dat )

mC.3b <- map2stan(
alist(
C ~ dbinom( 1 , Ip ) ,
logit(Ip) <- a + (bVC*VC + bVB*VB +bVT*VT) + bAVC*AVC,
a ~ dnorm(0, 10),
bVC ~ dnorm(0, 50),
bVB ~ dnorm(0, 10),
bVT ~ dnorm(0, 10),
bAVC~ dnorm(0, 10)
) ,
chains = 4,
data=dat )

mC3vec <- density( as.data.frame(mC.3@stanfit)[,"bVC"], adjust=2 )
mC3avec <- density( as.data.frame(mC.3a@stanfit)[,"bVC"], adjust=2 )
mC3bvec <- density( as.data.frame(mC.3b@stanfit)[,"bVC"], adjust=2 )

colors <- c(col.alpha("maroon",0.8), col=col.alpha("blue",0.8), col.alpha("darkgreen",0.8))
labels <- c("dnorm(-3, 1)", "dnorm(0, 10)", "dnorm(0, 50)")

matplot( mC3avec$x , mC3avec$y , type="l" , col=col.alpha("maroon",0.8) , pch=20 , lty="dashed",
         main="Comparison of bVC Posterior Distributions" , xlab="Param. Value" , ylab="Density"
)
lines( mC3vec$x , mC3vec$y , col=col.alpha("blue",0.8) , pch=20 , lty="dashed",
       main="" , xlab="Param. Value" , ylab="Density")
lines( mC3bvec$x , mC3bvec$y , col=col.alpha("darkgreen",0.8) , pch=20 , lty="dashed",
       main="" , xlab="Param. Value" , ylab="Density" )

legend("topright", inset=.05, title="Priors",
  labels, lwd=2, lty=c(1, 1, 1, 2), col=colors)

diagMCMC(attr(mC.3, "stanfit"), parName = c("bVC"))
diagMCMC(attr(mC.3a, "stanfit"), parName = c("bVC"))
diagMCMC(attr(mC.3b, "stanfit"), parName = c("bVC"))
```

The results could be viewed in figure 6. There were three lines representing posterior distribution for $\beta_{\{VC\}}$. A line showed the basic MCMC estimation, using $Normal(\mu = 0, \sigma = 10)$ in all estimations. Basically, $\sigma = 10$ showed there was no strong belief before estimation, at the same time, $\mu = 0$. The estimated coefficient was $\beta_{\{VC\}} = 3.11$.

When testing by lowering the strength of the belief through increasing $\sigma = 50$, MCMC estimation showed that the result was not much different; the two blue lines were almost completely overlapped. In contrast, when testing the model with a level of strong belief and the opposite assumption of denying altogether the coefficient, by lowering it to $\mu = -3$, and strongly regularizing the prior for the parameter of standard deviation, $\sigma = 1$, the estimate contracted to $\beta_{\{VC\}} = 2.55$.

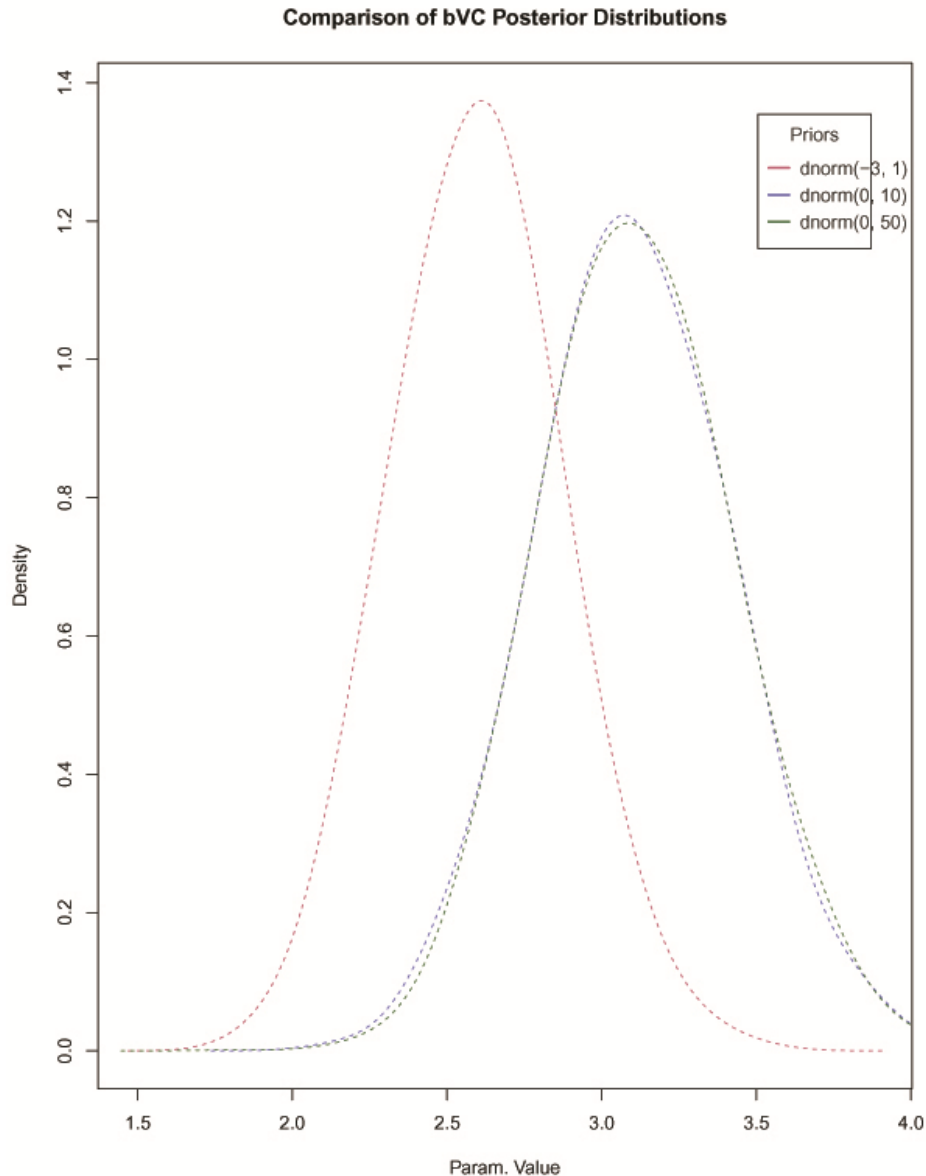

**Figure 6:** Prior testing by changing the values of parameters $\sigma$ and $\mu$.

In addition, we performed the MCMC technical validation for these estimations, with examinations being given in In figure 7. In general, all the technical properties met the requirements, even when the length of the Markov chain was not increased. Four posterior distributions showed a standard shape and almost completely overlapped (bottom right).

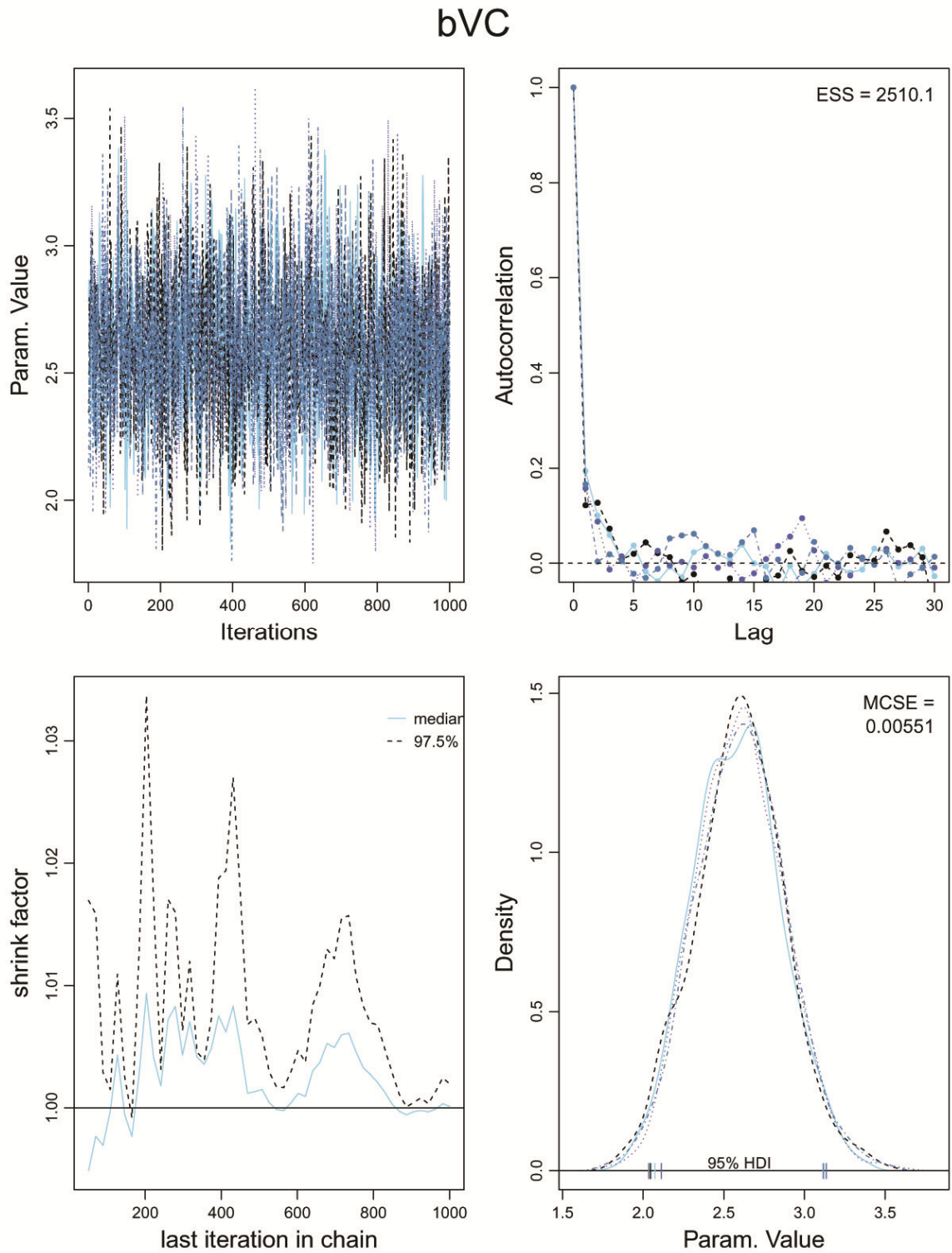

**Figure 7:** MCMC technical validation for the prior testing process.

The question was when translated into probability, was this level of reduction in estimate big or small? The answer was ready: logistic(2.55) = 92.8% compared with logistic(3.11) = 95.7%. In fact, the estimated standard deviation fluctuated around 0.3 even though in the prior-tests, the $\sigma$ was chosen quite "wildly": $\sigma = \{1, 10, 50\}$.

Other tests showed the same trend in the results. The big changes in priors did not lead to any considerable changes in the final results, which showed the model was highly robust. The open codes and data of all these estimations could be used to test the conclusions above.

**Analyzing the estimated models**

Firstly, we look at general comparisons for models of the same series in the following box. One of the main criteria for judging a preferred model for its goodness of fit is the "weight". (For a detailed explanation of this, refer to McElreath (2016).)

```
Comparing (mB.1), (mB.2), and (mB.3)
       WAIC pWAIC dWAIC weight     SE   dSE
mB.1  331.3   3.2   0.0   0.68  21.88    NA
mB.2  333.7   5.5   2.5   0.20  22.34  3.26
mB.3  334.7   5.1   3.4   0.12  22.23  1.57
Comparing (mC.1), (mC.2), and mC.3
       WAIC pWAIC dWAIC weight     SE   dSE
mC.1  322.0   3.6   0.0   0.44  23.09    NA
mC.3  322.1   5.7   0.1   0.41  23.49  4.41
mC.2  324.1   5.9   2.1   0.16  24.01  3.91
Comparing (mT.1), (mT.2), and (mT.3)
       WAIC pWAIC dWAIC weight     SE   dSE
mT.1  283.6   3.1   0.0   0.52  23.69    NA
mT.3  284.1   5.3   0.5   0.40  24.15  4.41
mT.2  287.3   5.7   3.7   0.08  24.45  3.68
```

Secondly, we would like to look at the models with "cultural additivity"; i.e. those with their names ended with 2 or 3. And their technical comparisons are given in the following box.

```
models T
       WAIC pWAIC dWAIC weight     SE   dSE
mT.3  284.1   5.3   0.0   0.83  24.15    NA
mT.2  287.3   5.7   3.2   0.17  24.45  5.78
models B
       WAIC pWAIC dWAIC weight     SE   dSE
mB.2  333.7   5.5   0.0   0.61  22.34    NA
mB.3  334.7   5.1   0.9   0.39  22.23  3.96
models C
       WAIC pWAIC dWAIC weight     SE   dSE
mC.3  322.1   5.7   0.0   0.72  23.49    NA
mC.2  324.1   5.9   1.9   0.28  24.01  3.24
```

Here, we used the logistic regression to consider the effects of "additivity" as a cultural phenomenon regarding religious faiths. In the models to calculate the typicality of Taoist stories, the strongest predictor was the model with only Taoist values, with mT1 takes 52% weight. However, we could not eliminate the simple additivity model (mT3) with 40% weight. The coefficients and MCMC statistical estimations of (mT3) presented below showed that the factor with the strongest influence was the coefficient of Taoist values, $\beta_{\{VT\}}$ = 3.96; next was the coefficient of anti-value of Taoism, $\beta_{\{AVT\}}$ = 2.44. They were the two purely Taoist elements.

Meanwhile, except for $a = -1.59$, then the additivity of $\beta_{\{VC\}}$ or the coefficient of Confucian values was also confirmed but with a negative sign (−). This meant the existence of Confucian values reduced the influence of Taoist values. The effect of Buddhist values could not be confirmed, because though $\beta_{\{VB\}}$ was negative (−), it moved widely to the two sides of zero.

| model (mT3) | Mean | StdDev | lower 0.89 | upper 0.89 | n_eff | Rhat |
|---|---|---|---|---|---|---|
| a | -1.59 | 0.23 | -1.98 | -1.25 | 3071 | 1 |
| bVC | -0.66 | 0.34 | -1.18 | -0.09 | 3419 | 1 |
| bVB | -0.16 | 0.39 | -0.77 | 0.47 | 3489 | 1 |
| bVT | 3.96 | 0.44 | 3.28 | 4.69 | 3377 | 1 |
| bAVT | 2.44 | 0.68 | 1.31 | 3.46 | 4000 | 1 |

In all three series B, C and T, it was only in model series B that (mB1) showed dominance. This meant (mB1) had the lowest additivity, with (mB1) takes 68% of weight, when compared with (mB2) and (mB3). However, if we didn't want to exclude the models of additivity, then the model worth analyzing was was (mB2) with 20% weight. If only (mB2) and (mB3) were compared, (mB2) was more dominant with 61% weight.

In the following box, the (mB2) model (complex additivity model of Buddhism), factors with the strongest influence were $\beta_{\{VB\}}$, and $\beta_{\{VAB\}}$. This meant that although the complex additivity model was more dominant, the influential factors were those with the Buddhist values.

| model (mB.2) | Mean | StdDev | lower 0.89 | upper 0.89 | n_eff | Rhat |
|---|---|---|---|---|---|---|
| a | -1.81 | 0.20 | -2.12 | -1.47 | 2617 | 1 |
| bVC | 0.69 | 0.54 | -0.19 | 1.55 | 2965 | 1 |
| bVB | 2.52 | 0.45 | 1.81 | 3.24 | 2471 | 1 |
| bVT | 0.44 | 0.63 | -0.59 | 1.38 | 3220 | 1 |
| bAVB | 0.91 | 0.31 | 0.39 | 1.40 | 2848 | 1 |

Finally, we considered the two most interesting models: (mC2) and (mC3). When comparing the weight of three models of Confucianism, the result was (mC1)/(mC2)/(mC3) = 44/40/16.

It was obvious that the appearances of (anti-) Confucian values (VC or AVC) should be associated with whether the key messages of a story were about Confucianism (C). Yet, after running the statistical test through 345 lines of data, it was quite surprising the model this logistical model with no additivity (mC1) was not much more weighted than the model (mC2) and (mC3), the ones with simple additivity.

This led us to prioritize additivity models in our analysis. Models (mC1) and (mC3) had the same weight, yet, the 16% weight of (mC2) suggested we should not ignore the potential influence and the explanatory strength of this model.

When considering only (mC3) with (mC2), the result was (mC3) takes 72% weight while it is 28% for the latter. This was not exactly strong enough evidence to eliminate the validity of (mC2). The following box showed the estimated coefficients of (mC2) and (mC3).

```
model (mC2)
       Mean StdDev lower 0.89 upper 0.89 n_eff Rhat
a     -2.03   0.25     -2.40      -1.62  1914    1
bVC    2.94   0.35      2.39       3.48  2040    1
bVB    0.01   0.43     -0.71       0.67  3511    1
bVT    0.86   0.57     -0.07       1.73  2915    1
bAVC   2.19   0.40      1.55       2.84  2463    1
model (mC3)
       Mean StdDev lower 0.89 upper 0.89 n_eff Rhat
a     -2.17   0.28     -2.61      -1.73  2067    1
bVC    3.11   0.31      2.63       3.62  2289    1
bVB   -0.01   0.34     -0.58       0.50  3167    1
bVT    0.73   0.37      0.10       1.29  3406    1
bAVC   2.19   0.40      1.57       2.85  2633    1
```

The strongest coefficients of the two models were $\beta_{\{VC\}}$ equaled 2.94 and 3.11 for (mC2) and (mC3), respectively. This was basically the same in terms of probability, as logistic (2.94) = 95% compared to logistic (3.11) = 95.7%. The second highest coefficient belonged to anti-Confucian-values elements, $\beta_{\{AVC\}}$ equaled 2.19 in both equations.

It is interesting to see that among all model series for the three religions, only the model series for Confucianism showed clearly the property of additivity, and both simple and complex models of additivity showed the existence of additivity is credible. Here, the appearance of Taoist elements (bVT) took on the coefficient values of 0.86 and 0.73.

Although technically speaking the simple additivity model (mC3) was more weighted, we might be more drawn toward the complex additivity model (mC2) as the co-existence and interaction with Confucian elements in the stories were stronger with $\beta_{\{VT\}} > 0$. The coefficient reflected the existence of predictor $VT \cdot VC$, through the equation:

$$C = -2.03 + 2.94VC + 0.86VT \cdot VC + 2.19AVC.$$

As can be seen from the equation, this interaction among elements of Confucianism and Taoism together increased the probability of categorizing a story into the $C$ category: it conveyed key messages about Confucianism.

On the one hand, this model reflected the logic in thinking about cultural influence and interaction, which had launched this statistical investigation. On the other hand, this was the only significant interaction model among all the estimated models. Figure 8 shows that the coefficient of $VT \cdot VC$, $\beta_{\{VT \cdot VC\}}$ satisfied the technical requirements similar to the case of $\beta_{\{VC\}}$ mentioned above.

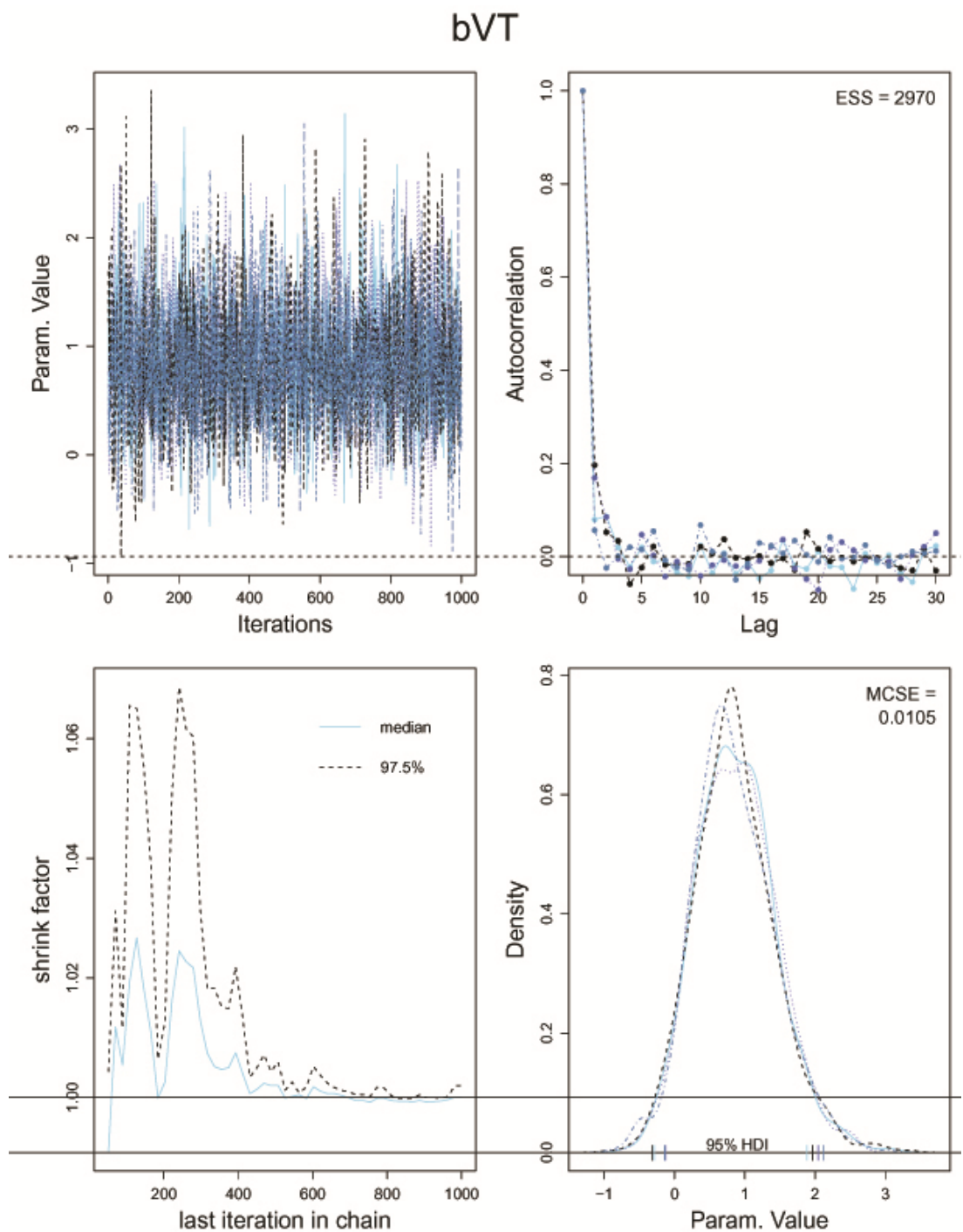

**Figure 8:** Hamiltonian MCMC technical validations for $\beta_{\{VT \cdot VC\}}$

**Discussions**

**Interpretation of the results**

Through the Bayesian categorical analysis, we have discovered and proven the existence of the cultural additivity phenomenon. Empirical results showed that the interaction or addition of Confucian and Taoist values in folklores together helped predict whether the key message of a story is Confucian.

For Buddhism, there was no such statistical confirmation. This finding was especially interesting because, as we have observed in the extant literature, it seemed that the two dominant religions in Vietnam are Confucianism and Buddhism. Up to this point, it has been implied that either all three religions Buddhism, Confucianism, and Taoism were on an equal footing (per the *Tam giáo đồng nguyên* principle), or Taoism was slightly dwarfed by the other two in terms of influence on society. However, empirical evidence suggested an isolation of Buddhism and higher possibility of additivity where it concerned Confucianism and Taoism.

One way to interpret this result was that cultural additivity (defined as the arbitrary tolerance of and willingness to add new beliefs, values or norms even when there was contradiction, to the existing system of belief) might be limited to cases where the two religions in interaction were not too far away or competing in terms of values. For example, Confucian and Taoist values seemed contradictory. Confucianism upheld a rigid hierarchy and equally rigid rules of conducts in different social context as well as *moral* obligations to study and become a member of the court, to be involved with the movement of the nation, of the society. The *wuwei* life of Taoism is almost the complete opposite: to retreat to isolation, to live closer to nature, to consider all concerns left behind as mundane, even though many of such concerns would have been obligations in Confucian principles. However, in a way, it was possible to adhere to both Confucianism and Taoism. Concretely speaking, this meant that after living the pragmatic life and restrictive path to virtue and honor prescribed by Confucianism, one could seek comfort in the isolation in nature offered by Taoism. In fact, one could even imagine that a person who has followed all this constricted life of Confucianism would, later in life, desire the Taoist ideals. A typical example might be Nguyễn Bỉnh Khiêm (1491-1585), the educator, poet, and sage, whose advice was sought after by many political leaders during his time, decided to retreat to isolation with nature later in life.

As for the isolation of Buddhism, the path of non-attachment, of "*phổ độ chúng sinh*" (to bring compassion to everyone), of self-cultivation with the practices of the Eight-fold way might be too demanding, thus the values and norms put forth by Buddhism might be too far away to be added together with the other two religions.

Another way to interpret the empirical results was that Confucianism, with all its pragmatic values, had more influence on Vietnamese culture than Buddhism and Taoism. In fact, equations in the previous section showed that even the Taoist elements in a story could enhance the likelihood that the story sent a message about Confucianism.

**Implications for the concept of Three Religions with common roots (*"Tam giáo đồng nguyên"*)**

Empirical results would show which of the meanings put forth to explain this concept was less relevant than others. Le (2016) and Nguyen (2015) had named several possible interpretations of *tam giáo đồng nguyên*: three religions share the same origin, the convergence of the three religions, the unification of three religions, the three religions are one, three religions co-exist in a cultural entity, etc.

The analysis had been focused on how values of different religions co-existed and interacted in Vietnamese folklore. For this reason, the results would not be able to show whether it was true that the three religions had the same origin, or the purposes of the three religions fully converged. However, the empirical evidence was definitively enough to prove that the three religions did not unify: Confucianism and Taoism showed a certain degree of additivity, while Buddhism did not.

The interpretation of *Tam giáo đồng nguyên* as religious unicity (the three religions could be considered as one, thus equal) was also more or less false. Regression equations showed that Confucianism was stronger than Taoism, so much so that when a story was comprised of both Confucian and Taoist values, the Confucian element could overpower the Taoist to the point that the key message of the story turned out predominantly Confucianism. And while it might be true to claim that the three religions co-existed in a cultural entity, the empirical results allowed us to make the case that there was more than just mere co-existence. There was indeed complex interaction and, seemingly, a domination of Confucianism over other religions.

**Implications for society**

The phenomenon of cultural additivity, defined as the readiness or the comfort in incorporating new beliefs, values and norms (even contradictory ones) into existing beliefs systems, had many important implications for Vietnamese society and its system of political power as well (Vuong & Vuong, 2018).

*Non-extremism*

First of all, the inherent tolerance in cultural additivity seemed to keep Vietnamese people away from religious sectarian conflicts, or any form of violent extremism. In general, when people divided themselves along the line of religious sects, some forms of violent conflicts over theological disputes were bound to happen. Notable conflicts included Protestants vs. Catholics in Northern Ireland or Sunni vs. Shia, Jews vs. Muslims in Palestine, Sinhalese Buddhists vs. Tamil Hindus in Sri Lanka (Harris, 2005). In contrast, there have been no "religious conflicts" in Vietnam according to Le (2011). A quick search on Google Scholar would show that there were no results for the key words "religious wars in Vietnam". Religions co-existed quite harmoniously in Vietnam. In the 19th century, Vietnamese people had even invented a syncretic religion called Caodaiism *("Đạo Cao Đài")*, whose essence was essentially additivity: founded in 1926, this faith brings together elements of the East and the west by worshipping Buddha, Lao Tzu, Victor Hugo, Confucius, Jesus Christ, etc. (Hitchcox, 1990; Quinn‑Judge, 2013), and even considered the Jade Emperor (*Ngọc Hoàng*) and the Holy Spirit to be one (Louaas, 2015; Hoskins, 2015).

*Lack of critical thinking, thus creativity, in education, science, and technology*

Second, this seemingly extreme tolerance and compromise, however, might not be positive for critical thinking. In fact, it might even result in a certain level of outrageous sophistry. For example, in Vietnamese media during November 2017, there has been a heated debate over whether a doctoral graduate of a famous graduate school has plagiarized or not. Though the person had copied paragraphs from a textbook to add to her dissertation without citation and reference to the textbook, there were still professors who went on records and defended this freshly-minted Ph.D. They argued that the behavior was not plagiarism in the sense that it had no ethically reprehensible intent; to them, it was rather an incidence of "unprofessional referencing" or "technical mistakes" (Dang, 2018). (Their comments soon became a joke in the whole Vietnam.) This lack of critical thinking, viewed under the lenses of cultural additivity, might be the result of an extreme case of how ready and comfortable people were arbitrarily incorporating new "ideas" into their own, as well as compromising on the basic principles (including ethics).

In this extreme example, we could observe that the elite of the education system was unable of making a principled judgment with regards to plagiarism. This was only a symptom of the weakness of the education system of Vietnam in promoting introspective skills such as self-questioning or self-examination. This lack of autonomy in thinking occurred from grassroots level: the average grade student, when faced with problems that required creative or critical thinking, would be inclined not to find ways to eliminate bad options but to find solutions through adding more elements from any sources they could find.

The negative dragging effects of cultural additivity, thus, might have extended to science and technology, as they were areas that required perhaps the highest level of creative problem-solving skill. Indeed, though Vietnam's scientific output is steadily rising, it is still low compared with countries in ASEAN (Ho et al., 2017; Nguyen & Pham, 2011); and a recent study found that 75% researchers in Vietnamese social sciences have never attempted single-authored paper in the past 10 years (Vuong et al., 2017).

*Misleading branding practice in business*

Third, in the field of business, this study about the case of Vietnam supplemented a previous research on cultural differences in resolving information incongruity in North American and East Asian cultures. Particularly, when looking at the way individuals processed information incongruity, Aaker and Sengupta (2000) found that those in a North American culture tend to adopt the attenuation strategy, in which evaluations were affected by the "more diagnostic attribute information", therefore sought to resolve incongruity. On the contrary, individuals of East Asian culture were often less compelled to resolve the incongruity for they would follow an additivity strategy wherein there is the incorporation of even conflicting pieces of information (Aaker & Sengupta, 2000).

Vietnamese people, as our study has shown, were inclined to arbitrarily add new ideas, even conflicting ones, to their existing belief system. In the language of information incongruity, the Vietnamese, like members of other East Asian cultures, would also adopt the additivity strategy when faced with two opposing pieces of information. What this behavior entailed in the business sphere was, among other things, branding and sales tactics that turned out to be somewhat manipulative. For instance, in Vietnam, advertisements for food supplements often branded these products as medicines even though they were not medicines by any medical or legal definition. By doing so, producers and sellers of food supplements hit the target consumer group—those who wanted to cure a certain ailment without resorting to drugs. Yet, at the same time, the misleading advertisements have also resulted in confusion, and in the long run, possibly ineffectiveness of treatment.

## Conclusion

While observations of the Vietnamese's openness to any kind of religious and spiritual activity that met their daily needs were made by many scholars or that "Vietnamese people had the ability to integrate new traditions into their existing beliefs systems, blending traditions, together without seeing them as contradictory" (Tran, 2018, pp. 15), statistical evidence to describe the nature, the mechanism and results of this ability was clearly lacking.

In this study, through the application of Bayesian categorical analysis (Stan/Hamiltonian MCMC in the R environment), we were able to statistically verify the existence of the widespread and long-standing social phenomenon defined as "cultural additivity": the nature of the interaction and co-existence of contradictory the values, norms and beliefs of the *Three Religions* in Vietnamese long-standing folktales. Concretely speaking, empirical results showed the interaction or addition of Confucian and Taoist values in folklores together help predict whether the key message of a story was Confucian, while there is no such statistical tendency for Buddhism.

As the pattern of cultural additivity among the three religions' beliefs, values and norms showed a certain level of domination of Confucianism, there were many implications for society. The first important one was how religion, despite having been named the explicit cause of various conflicts throughout the world, has not given rise to extremist behaviors in Vietnam: there had been nearly no religious conflict or extremism in Vietnamese history. Second, in terms of critical thinking, the cultural additivity found here might be attributable to the high degree of compromise in original ideas, and even to some extent outrageous sophistry. This is because in extreme cases, some people might be overly willing to appropriate new ideas of others as their own and reluctant to create anything new. What this entailed was an education system poorly prepared to form a principled judgment for plagiarism and a science and technology sector lagging far behind that of regional countries.

What is notable in this study is the efficiency of Bayesian statistics in quantitatively analyzing a cultural phenomenon ("cultural additivity") using traditionally qualitative materials, i.e. Vietnamese folkloric literature. This method opened up a potential direction for all quantitative research on culture, religion, ethics, sociology, among others; potentially, this direction could supplement the shortcomings that research on human neurons and brains often faced—such as making claims at the societal level. In addition, given its explanatory power, the term "cultural additivity" used in this study could indeed benefit from more research in the future to get a more comprehensive account.

The cultural additivity as seen among the Vietnamese people, after all, might be the reason behind the country's flexible adoption of, and adaptation to, new ideas, be it religions or languages. Perhaps this was also why in Vietnam, one would see people burning joss papers in the shape of iPhones and airplanes even if they neither knew the origin of the practice nor knew if iPhones and airplanes were compatible with the spirits that they were trying to appease, but one would never find the scene of an elevator boy specialized in pushing the buttons on behalf of the Jewish people during Shabbat[2], as bemused by the physicist Richard Feynman in an encounter with the Orthodox Jews.

> "Here [the rabbinical students] are, slowly coming to life, only to better interpret the Talmud. Imagine! In modern times like this, guys are studying to go into society and do something—to be a rabbi—and the only way they think that science might be interesting is because their ancient, provincial, medieval problems are being confounded slightly by some new phenomena." — Feynman (1985).

For closing these important socio-cultural remarks after lengthy discussions on Vietnamese beliefs, values, and ideals, we look at the linocut painting "Days near *Tết*" in Figure 8 by Bui Quang

---

[2] The rabbinical students did not push the elevator buttons out of the fear that the electric spark was like fire, which the Jews are not allowed to touch during Shabbat. Feynman assured them that "Electricity is not fire. It's not a chemical process, as fire is". (Feynman, 1985)

Khiem, in which some old women, perhaps very old, sold joss papers and other cultural items nearby a tree root. (For more information about *Tết* holiday in Vietnam, see Pham (2018).)

**Figure 8:** Days near *Tết*

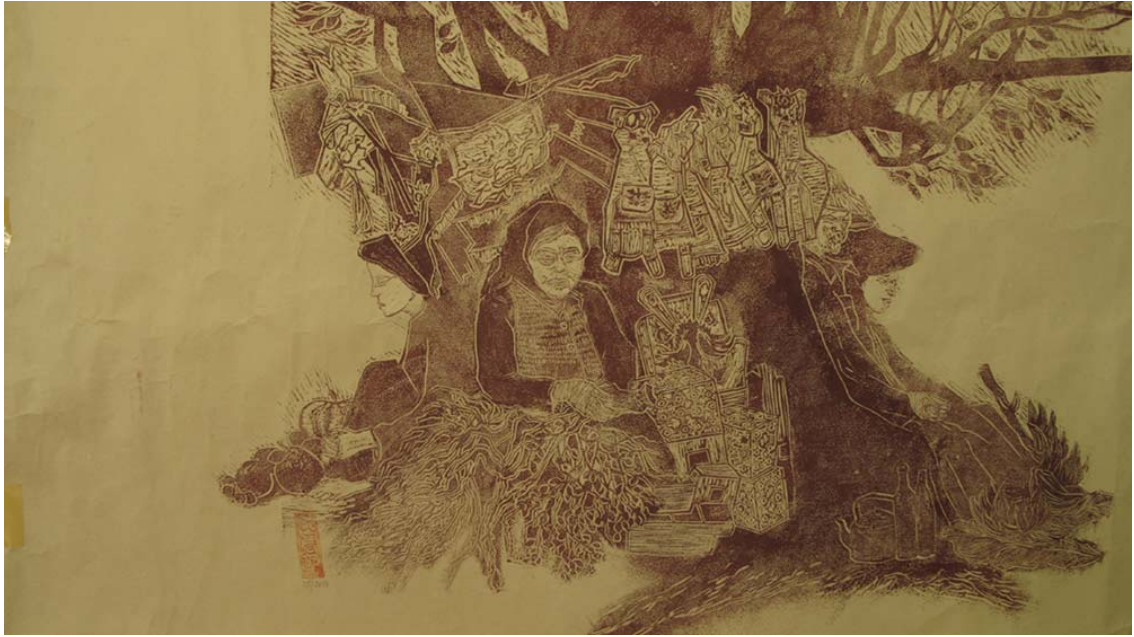

The painting is presented here for a reason. Vietnamese people added numerous things and meanings to their livelihood, places, and beliefs. The women themselves were then added and became part of the root itself. Perhaps now the root has its soul and is worshipped by numerous people who live nearby. And Vietnamese have for long believed that "*Thần cây đa, ma cây gạo*" still holds true (literally, the banyan trees have God, Bombax ceiba have Ghost inside).

**Acknowledgment:**

We would like to extend our thanks to Dam Thu Ha, of Vuong & Associates, for her assistance during the research process.